\documentclass[ conference,a4paper]{IEEEtran} 

\usepackage{amssymb}
\usepackage{amsmath}
\usepackage{amsfonts}
\usepackage{graphicx}
\usepackage{algorithm}
\usepackage{algorithmic}
\usepackage{epstopdf}
\usepackage{cite}
\usepackage{stfloats}
\usepackage{url}
\usepackage{CJK}
\usepackage{hyperref}
  \usepackage{epstopdf}
  \usepackage{subfigure}
\usepackage{algorithm,algorithmic}
\usepackage{multicol}

\usepackage{color,xcolor}

\newtheorem{theorem}{\text{Theorem}}

\renewcommand\thesubsection{\Alph{subsection}} 
\newtheorem{Prob}{Problem}

\IEEEoverridecommandlockouts

\begin{document}
\title{Joint Optimization of Preamble Selection and Access Barring for MTC with Correlated Device Activities }

%

\author{\IEEEauthorblockN{Wang Liu, Ying Cui, Lianghui Ding, Jun Sun, Yangyang Liu, Yang Li, Li Zhang\thanks{W. Liu, Y. Cui, L. Ding and J. Sun are with the Department of Electronic Engineering, Shanghai Jiao Tong University, China. Y. Liu is with Shanghai Electro-mechanical Engineering Institute, China. Y. Li and L. Zhang are with Wuhan Maritime Communication Research Institute, China.}}}



\maketitle

\begin{abstract}
Most existing works on random access for machine-type communication (MTC) assume independent device activities. However, in several Internet-of-Things (IoT) applications, device activities are driven by events and hence may be correlated.
This paper investigates the joint optimization of
preamble selection and access barring for correlated device
activities.
We adopt a random access
scheme with general random preamble selection parameterized
by the preamble selection distributions of all devices and an access barring scheme parameterized
by the access barring factor, to
maximally exploit correlated device activities for improving the
average throughput.
 First, we formulate the average throughput maximization problem with respect to the preamble selection distributions and the access barring factor. It is a challenging nonconvex problem.
 We characterize an optimality property of the problem.
 Then, we develop two iterative algorithms to obtain a stationary point and a \text{low-complexity} solution respectively by using the block coordinate descend (BCD) method.
Numerical results show that the two proposed solutions achieve significant gains over existing schemes, demonstrating the significance of exploiting correlation of device activities in improving the average throughput.
 Numerical results also show that compared to the stationary point,
 the low-complexity solution achieves a similar average throughput with much lower computational complexity, demonstrating the effectiveness of the low-complexity solution.
\end{abstract}


\section{Introduction}

%
Internet-of-Things (IoT) has had broad applications in several areas, such as home automation, smart grids, healthcare systems, and industrial monitoring, and has received increasing attention in recent years. The number of IoT devices is expected to grow up to 30 billion by 2030, and more and more new IoT applications are emerging. There is a need to design a robust, scalable, and efficient sixth-generation (6G) wireless network that can effectively realize machine-type communications (MTC) to support future IoT applications. This paper aims to provide a promising solution for random access for MTC in 6G.


In random access for MTC, devices compete in a random access channel (RACH) to access a base station (BS) through the random access procedure~\cite{TR_3gppevolved}.
Specifically, each active device randomly selects a preamble from a pool of available preambles according to a preamble selection distribution and transmits it during the RACH.
The BS acknowledges the successful reception of a preamble if such preamble is transmitted by only one device.
In\text{\cite{Wong16TVT, Wong15TWC,  Kellerer19TWL ,Dailin18TWC,  QL_IOTJ2019, Seo_TVT2017,Lee2015TWC, Choi_CL2016 }}, the authors consider the random access procedure and study the effect of preamble selection under certain assumptions on the knowledge of device activities.
Specifically,\text{\cite{Wong16TVT, Wong15TWC,  Kellerer19TWL , QL_IOTJ2019,Lee2015TWC, Choi_CL2016 }} assume that the number of active devices is known;~\cite{ Seo_TVT2017 } assume that the distribution of the number of active devices is known;~\cite{Dailin18TWC} assume that the statistics of the data queue of each device are known.
In\text{\cite{Wong16TVT, Wong15TWC,  Kellerer19TWL ,Dailin18TWC, QL_IOTJ2019, Seo_TVT2017,Lee2015TWC, Choi_CL2016 }}, preambles are selected according to a uniform distribution, and the average throughput~\cite{Wong16TVT, Wong15TWC,  Kellerer19TWL ,Dailin18TWC, QL_IOTJ2019, Seo_TVT2017,Lee2015TWC, Choi_CL2016},  average access delay~\cite{Lee2015TWC} and resource consumption~\cite{ Kellerer19TWL} are analyzed.
Assuming that all devices have the same preamble selection distribution, the authors in~\cite{ Kellerer19TWL ,Lee2015TWC, Choi_CL2016 } optimize the number of allocated preambles to maximize the average throughput~\cite{Kellerer19TWL,Choi_CL2016 } or access efficiency~\cite{Lee2015TWC}.

When many devices attempt to access a BS simultaneously, a preamble is very likely to be selected by more than one device, and hence the probability of access success decreases significantly.
In this scenario, access control is necessary.
One widely used access control method is the access barring scheme, which has been included in the LTE specification in~\cite{TR_3gppevolved}.
In\text{\cite{Wong16TVT, Wong15TWC,  Kellerer19TWL ,Dailin18TWC,  QL_IOTJ2019, Seo_TVT2017,Lee2015TWC}}, the authors also consider access barring.
Specifically, the access barring factor is optimized to maximize the average throughput~\cite{ Wong16TVT, Wong15TWC,  Kellerer19TWL ,Dailin18TWC,  QL_IOTJ2019, Seo_TVT2017} or access efficiency~\cite{Lee2015TWC}.

In \cite{Dailin18TWC,  Wong16TVT,   Kellerer19TWL , QL_IOTJ2019, Seo_TVT2017,Lee2015TWC,  Wong15TWC , Choi_CL2016 }, the activities of a set of devices are assumed to be independent and identically distributed (i.i.d).
However, in many IoT applications, such as smart metering and environment sensing, device activities are driven by events and are hence correlated.
The preamble selection distributions and access barring factors designed for i.i.d device activities may not be effective for devices with correlated activities.
To our knowledge, \cite{Popovski18SPAWC} is the first work that considers general (possibly correlated) device activities and the optimization of the preamble selection distributions and access barring factors of all devices under a general device activity distribution.
More specifically, in \cite{Popovski18SPAWC}, the authors maximize an approximation of the average throughput which captures the activity probabilities of a single device and every two devices and develop a heuristic algorithm to tackle the challenging nonconvex problem.
The approximation error and the heuristic algorithm may yield a non-negligible loss in the average throughput.
Therefore, it is critical to explore more effective algorithms for the case of correlated device activities.


This paper considers MTC with correlated device activities, which plays a key role for future IoT and 6G.
We adopt a random access
scheme with general random preamble selection parameterized
by the preamble selection distributions of all devices and an access barring scheme parameterized
by the access barring factor, to
maximally exploit correlated device activities for improving the
average throughput.
First, we formulate the average throughput maximization problem, which is a nonconvex problem with a complicated objective function.
We characterize an optimality property of the problem.
Then, based on the block coordinate descend (BCD) method, we develop an iterative algorithm to obtain a stationary point of the original problem which utilizes the correlation of the activities of all devices.
Next, using BCD, we also develop a low-complexity iterative algorithm to obtain a stationary point of an approximate problem which exploits the correlation of the activities of every two devices.
Finally, by numerical results, we show that the proposed solutions achieve significant gains over existing schemes, demonstrating the significance of exploiting correlation of device activities in improving the average throughput; and we show that the average throughput of the low-complexity solution is close to that of the stationary point, demonstrating that exploiting the
	correlation of the activities of every two devices in a rigorous
	way is almost sufficient.
The key notation used in this paper is listed in Table I.

\section{System Model }
 \begin{table}[t]
		\caption{Key Notation}
	\centering
	\begin{tabular}{|c|c|}\hline
		Notation&Description \\ \hline
		$K$& the number of devices \\ \hline
			$N$&  the number of preambles\\ \hline
				$x_k \in \{0,1\}$&  the activity state of device $k$\\ \hline
		$\bf x$& the activity states of all $K$ devices\\ \hline
		$ p_{\bf x}$& the probability that the activity states are $\bf x$ \\\hline
		$\bf p$& the general activity distribution\\ \hline
		$\epsilon \in [0,1]$	& the access barring factor \\\hline
		$a_{k,n}\in [0,1]$  & the probability that device $k $ selects preamble $n$ \\\hline
		$\bf A$  & the preamble selection distributions of all $K$ devices \\\hline
		$ 	T({\bf A},{\epsilon},{\bf x})$	& average throughput conditional on $\bf x$    \\\hline
		$ \bar	T({\bf A},{\epsilon},{\bf p})$	& average throughput   \\\hline
		$ \tilde	T_{\lambda}({\bf A},{\epsilon},{\bf p})$	& approximate average throughput parameterized by $\lambda$   \\\hline
	\end{tabular}
	\label{tab:Margin_settings}
\end{table}

\begin{figure}[t]
	\centering
	\includegraphics[width=0.30\textwidth]{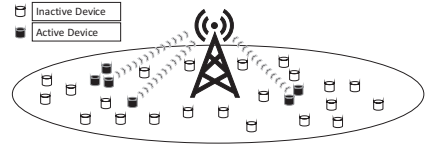}
	\vspace{-1mm}
	\caption{\small{System model.}}\label{fig_SystemModel}
	\vspace{-5mm}
\end{figure}

We consider the uplink of a single-cell wireless network consisting of
one BS and {$K$} MTC devices, as shown in Fig.~\ref{fig_SystemModel}.
Let $\mathcal K \triangleq \{1,2,...,K\}$ denote the set of $K$ devices.
Consider a discrete-time system with time being slotted.
In practical MTC, adjacent devices may be triggered by the same event or dependent events.
Thus, we assume that within each time slot, the devices activate according to a general $K$-dimensional binary distribution.
{It is worth noting that in} most previous work \cite{Wong16TVT, Wong15TWC,  Kellerer19TWL ,Dailin18TWC,  QL_IOTJ2019, Seo_TVT2017,Lee2015TWC,  Choi_CL2016}, correlation of device activities is not considered, i.e., devices are assumed to activate independently.
Let
$x_k \in \{0,1\}$ denote the activity state of device $k$,
where $x_k=1$ if device $k$ is active, and $x_k=0$ otherwise.
Let ${\bf x}\triangleq [ x_1,x_2,...,x_K ]^T$ denote the activity states of all $K$ devices.
The device activity distribution is denoted by ${\bf p}\triangleq (p_{\bf x})_{{\bf x}\in \mathcal X} $, where $p_{\bf x}$ represents the probability of the activity states of the $K$ devices being $\bf x$ and $  \mathcal X \triangleq \{0,1\}^K$.
Note that
\begin{subequations}\label{C_DAP}
	\begin{align}
		&   0\leq p_{\bf x} \leq1, \  {\bf x} \in \mathcal X, \label{Cp1} \\
		& \sum\limits_{{\bf x} \in \mathcal X} p_{\bf x} =1.  \label{Cp2}
	\end{align}
\end{subequations}
In practice, $ p_{\bf 0} < 1 $.

In a slot, each active device tries to access the BS.
	Congestion may occur when a massive number of active devices
	require to access the BS at the same time.
	We adopt access barring scheme for access control~\cite{Wong16TVT}.
	In particular, at the beginning of each time slot, each active device
	independently
	attempts to access the BS with probability $\epsilon$, where
	\begin{subequations}\label{C_ACB}
		\begin{align}
			&\epsilon \geq 0, \label{C1} \\
			&\epsilon \leq1.  \label{C12}
		\end{align}
	\end{subequations}
	Here, $\epsilon$ is referred to as the access barring factor and will be optimized later.
	That is, the access barring scheme is parameterized by the access barring factor $\epsilon$.

\begin{figure*}[t]
	\newcounter{TempEqCnt} 
	\setcounter{TempEqCnt}{\value{equation}} 
	\setcounter{equation}{7} 
	
	\begin{align}
			&\frac{\partial \bar T ({\bf a}_k, {\bf a}_{-k},\epsilon,{\bf p})}{ \partial a_{k,n} }
			=\sum\limits_{m=1}^{K}
			-m(-\epsilon)^{m}  \!\!\!\!
			\sum\limits_{
				\mathcal K' \subseteq {\mathcal K}: k\in \mathcal K', |\mathcal K' |=m}
			\left( \sum\limits_{{\bf x \in \mathcal X}}
			p_{\bf x}
			\prod\limits_{l \in \mathcal K'  }{x}_{l}\right) \prod\limits_{l \in \mathcal K' : l\neq k  }
			\!\!\!\! a_{l,n} \triangleq Q_{k,n}({\bf a}_{-k}, \epsilon,{\bf p})
			, \quad  n\in\mathcal N,  k\in\mathcal K,  \label{Q_{k,n}}    \\
			&\frac{\partial \bar T ({\bf A}, \epsilon,{\bf p})}{\partial \epsilon}
			=
			\sum\limits_{m=1}^{K}
			m^2(-\epsilon)^{m-1}\sum\limits_{n\in \mathcal N}\sum\limits_{
				\mathcal K' \subseteq {\mathcal K}:  |\mathcal K' |=m}
			\left( \sum\limits_{{\bf x \in \mathcal X}}
			p_{\bf x}
			\prod\limits_{k \in \mathcal K'}{x}_{k}\right)  \prod\limits_{k \in \mathcal K'}  a_{k,n}  \triangleq q({\bf A},\epsilon,{\bf p}). \label{q}
	\end{align}
	\hrulefill
\end{figure*}
We adopt the random access procedure which consists of four stages, i.e., \textit{preamble transmission}, \textit{random access response}, \textit{scheduled transmission} and \textit{contention resolution}~\cite{TR_3gppevolved}.
We focus only on the first stage, where the success of access is mainly determined \cite{Wong16TVT, Wong15TWC,  Dailin18TWC,  QL_IOTJ2019, Seo_TVT2017,Lee2015TWC,  Choi_CL2016}.
Consider $N$ orthogonal preambles, the set of which is denoted by $\mathcal N \triangleq \{1,2,...,N\}$.
Specifically, at the first stage, each device that attempts to access the BS independently selects a preamble out of the $N$ preambles to transmit.
The probability that device $k$ selects preamble $n$ is denoted by $a_{k,n}$,
which satisfies
\begin{subequations}\label{C_PSDs}
	\begin{align}
		&\sum\limits_{n \in \mathcal N} a_{k,n}= 1, \  k\in \mathcal K, \label{C2} \\
		& \ a_{k,n} \geq 0, \ k \in \mathcal K, n \in \mathcal N. \label{C3}
	\end{align}
\end{subequations}
Let ${\bf a}_{k} \triangleq  (a_{k,n})_{  n \in \mathcal N}$
denote the preamble selection distribution of device $k$.
Let ${\bf A}\triangleq ({ a}_{k,n})_{k \in \mathcal K, n\in \mathcal N}$ denote the distributions of the $K$ devices.
The $k$-th row of $\bf A$ is ${\bf a}_k$.
Note that for all $k \in \mathcal K$, the random \textit{preamble transmission} parameterized by ${\bf a} _k$ reduces to the \textit{preamble transmission} in the standard random access procedure \cite{TR_3gppevolved} when $a_{k,n} =\frac {1}{N}$, $ n \in \mathcal N$.
Furthermore, note that for all $k \in \mathcal K$, the considered preamble selection is in general random, and becomes deterministic when $a_{k,n} \in \{0,1\}, n \in \mathcal N$.
We allow ${\bf a}_{k}$, $k \in \mathcal K$ to be arbitrary distributions to more effectively avoid collision caused by the correlation of device activities.

If a preamble is selected by a single device, this device successfully accesses the BS~\cite{Dailin18TWC}.
Then, the average number of devices that successfully access the BS at activity states $\bf x$ in a slot is given by \cite{Popovski18SPAWC}
\begin{align}
	T({\bf A},{\epsilon},{\bf x})  \triangleq \sum\limits_{n\in \mathcal N}
	\sum\limits_{k\in \mathcal K} x_k a_{k,n} \epsilon \prod\limits_{l\in {\mathcal K:l \neq k}} (1-x_l a_{l,n}\epsilon ), \label{T_x}
\end{align}
where the average is taken over random access control and random preamble selections.

In this paper, to obtain first-order design insights, we assume that the device activity distribution has been estimated by some learning methods as in~\cite{Popovski18SPAWC}, and the estimation error is negligible. That is, the exact value of $\bf p$ is known.
We adopt the average throughput \cite{Popovski18SPAWC}
\begin{align}
		\setcounter{equation}{4}
&	\bar T({\bf A},{\epsilon}, {\bf p} )=  \sum\limits_{{\bf x} \in \mathcal X}p_{\bf x}    T({\bf A},{\epsilon},{\bf x})  \label{T_Avg.New}
\end{align}
as the performance metric, where $ T({\bf A},{\epsilon},{\bf x})$ is given by (\ref{T_x}).




\section{Problem Formulation}
In this section, we formulate the average throughput maximization problem which
 is a challenging non-convex problem,
 and characterize an optimality property of the optimal point.
Specifically, we optimize the preamble selection distributions $\bf A$ and the access barring factor $\epsilon$ to maximize the average throughput $ \bar T({\bf A},{\epsilon}, \bf p )$ in (\ref{T_Avg.New}) subject to the constraints on $({\bf A},\epsilon)$ in (\ref{C1}), (\ref{C12}), (\ref{C2}) and (\ref{C3}).
\begin{Prob}[Average Throughput Maximization]\label{Prob:Perfectcase}
	\begin{align}
		&\max_{\bf{A},\epsilon} \quad \bar T(\mathbf {A},\epsilon,\bf p)  \nonumber\\
		&\ \mathrm{s.t.}
		\quad \text{(\ref{C1})}, \ \text{(\ref{C12})}, \ \text{(\ref{C2})},\  \text{(\ref{C3})}. \nonumber
	\end{align}
\end{Prob}

 Note that the objective function $\bar T(\mathbf {A},\epsilon,\bf p)$ of Problem~\ref{Prob:Perfectcase}, which is the exact average throughput, reflects the correlation of the activities of all devices.
 In contrast, \cite{Popovski18SPAWC} optimizes two approximate functions of the average throughput which only reflect the correlation of the activities of every two devices.
 Intuitively, solving Problem~\ref{Prob:Perfectcase} will provide random access design that is more applicable to IoT applications with correlated device activities.
 The objective function $\bar T(\mathbf {A},\epsilon,\bf p)$ is nonconcave in $({\bf A},\epsilon)$, and the constraints in (\ref{C1}), (\ref{C12}), (\ref{C2}) and (\ref{C3}) are linear.
 Thus, Problem~\ref{Prob:Perfectcase} is nonconvex.
 In general, a globally optimal point of a nonconvex problem cannot be obtained
 effectively and efficiently.
 Obtaining a stationary point is the classic goal for dealing with a nonconvex problem.
 However, we can characterize an optimality property of a globally optimal point of Problem~\ref{Prob:Perfectcase}.

\begin{theorem}[Optimality Property] \label{Lem_Prop_Opt}
There exists at least one globally optimal point $({\bf A}^{*},\epsilon^{*})$ of Problem~\ref{Prob:Perfectcase} which satisfies
	${\bf a}_{k}^{*} ={\bf e}_{n_k},  k\in \mathcal K$,
	where ${\bf e}_{n_k}$ is an $N$-dimensional vector of all zeros except
	the $n_k$-th entry being $1$.
\end{theorem}

\begin{IEEEproof}
	Denote the algorithm mapping defined by steps 4 to 10 in Algorithm \ref{Alg_PBCD} that sends $\left({\bf A}^{(i)},\epsilon^{(i)}\right)$ in iteration $i$ to $\left({\bf A}^{(i+1)},\epsilon^{(i+1)}\right)$ in iteration $i+1$ by $f:\mathbb{R}^{KN+1}\rightarrow \mathbb{R}^{KN+1}$. The idea of the proof is to show that $({\bf A}^{*},\epsilon^{*}) \triangleq f({\bf A}^{\dag},\epsilon^{\dag})  $ is an optimal point satisfying the optimality property in Theorem~\ref{Lem_Prop_Opt}, where $({\bf A}^{\dag},\epsilon^{\dag})$ is an arbitrary optimal point.
\end{IEEEproof}

Theorem \ref{Lem_Prop_Opt} indicates that there exists a deterministic preamble
 selection rule that can achieve the maximum average throughput.
In Section IV and Section V, we shall see that the proposed stationary point and
 low-complexity solution satisfy the optimality property in Theorem~\ref{Lem_Prop_Opt}.

\begin{figure*}[t]
	\newcounter{TempEqCnttildeT} 
	\setcounter{TempEqCnttildeT}{\value{equation}} 
	\setcounter{equation}{11} 
	
\begin{align}
		\tilde T_{\lambda}(\mathbf {A},\epsilon,{\bf p}) & \triangleq
		\epsilon  \sum\limits_{n\in \mathcal {N}}\sum\limits_{k\in \mathcal {K}} a_{k,n}\sum\limits_{{\bf x} \in \mathcal X }p_{\bf x}  x_k
		-2\lambda\epsilon^2 \sum\limits_{n\in \mathcal {N}} \sum\limits_{k\in \mathcal {K }}a_{k,n} \sum\limits_{l \in {\mathcal K}: l >k}a_{l,n}
		\sum\limits_{{\bf x} \in \mathcal X }p_{\bf x} x_kx_l ,
		\nonumber \\
		&=   \epsilon \sum\limits_{k\in \mathcal {K}} \sum\limits_{{\bf x} \in \mathcal X }p_{\bf x}x_k
		-2\lambda\epsilon^2 \sum\limits_{n\in \mathcal {N}} \sum\limits_{k\in \mathcal {K }}a_{k,n} \sum\limits_{l \in {\mathcal K}: l >k}a_{l,n}
		\sum\limits_{{\bf x} \in \mathcal X }p_{\bf x} x_kx_l .
		\label{T_LB}
	\end{align}
	\hrulefill
\end{figure*}

\section{Stationary Point }
Based on the BCD method, we propose an iterative algorithm to obtain a stationary point of Problem~\ref{Prob:Perfectcase}.
Specifically, we divide the variables $\left({\bf A}, \epsilon\right)$
into $K+1$ blocks, i.e., ${\bf a}_{k}, \ k \in \mathcal K $ and $ \epsilon $.
In each iteration of the proposed algorithm, all $K+1$ blocks are sequentially updated once.
At each step of one iteration,
we maximize $  \bar T(\mathbf {A},\epsilon,\bf p)  $
with respect to one of the block.
{For ease of illustration, in the following, we also write
	$\bar T(\mathbf {A},\epsilon,\bf p)$ as $\bar T( {\bf a}_{k},{\bf a}_{-k} ,\epsilon,\bf p)$, where ${\bf a}_{-k}\triangleq({\bf a}_j)_{j\in {\mathcal K}, j\neq k} $.}
Given ${\bf a}_{-k}$ and $\epsilon$ obtained in the previous step, the
block coordinate optimization with respect to
${\bf a}_{k}$ is given by
\begin{align}
		\setcounter{equation}{5}
	&\max_{{\bf a}_k} \quad
	\bar T\left( {\bf a}_k ,{\bf a}_{-k} ,  \epsilon,\bf p \right) , \quad  k\in \mathcal K\label{Prob:Perfectcase_RAO}\\
	&\ \mathrm{s.t.}   \quad   \text{(\ref{C2})}, \ \text{(\ref{C3})}. \nonumber
\end{align}
Given $\bf A$ obtained in the previous step, the block coordinate optimization with respect to $\epsilon$ is given by
\begin{align}
	&\max_{\epsilon} \quad
	\bar T\left({\bf A}, \epsilon ,\bf p \right)
	\label{Prob:Perfectcase_ACB}  \\
	&  \ \mathrm{s.t.}  \quad   \text{(\ref{C1})}, \ \text{(\ref{C12})}.   \nonumber
\end{align}
Each problem in (\ref{Prob:Perfectcase_RAO}) is a linear program (LP) with $N$
variables and $N+1$ constraints.
The problem in (\ref{Prob:Perfectcase_ACB}) is a polynomial programming
with a single variable and two constraints.
Next, we obtain optimal points of the problems in (\ref{Prob:Perfectcase_RAO}) and (\ref{Prob:Perfectcase_ACB}).

Let $Q_{k,n}({\bf a}_{-k}, \epsilon,{\bf p}), k\in \mathcal K, n\in \mathcal N$ and $q({\bf A},\epsilon,{\bf p})$ denote the partial derivatives of $ \bar T ({\bf a},\epsilon, \bf p)$ with respect to $a_{k,n}, k\in\mathcal K, n\in \mathcal N$ and
$\epsilon$, respectively, as shown in (\ref{Q_{k,n}}) and (\ref{q}) at the top of this page.
Denote $\mathcal B({\bf A},{\bf p}) \triangleq \{z \in [0,1]: q( {\bf A},z,{\bf p}) =0\}$ as the set of roots of equation
$ q({\bf A},z,{\bf p})=0$ that lie in interval $[0,1]$.
Based on structural properties of the block coordinate optimization
problems in (\ref{Prob:Perfectcase_RAO}) and (\ref{Prob:Perfectcase_ACB}), we can
obtain their optimal points.
\begin{theorem}[Optimal Points of Problems
	in (\ref{Prob:Perfectcase_RAO}) and (\ref{Prob:Perfectcase_ACB})]\label{Thm_OptimmalSolutionBCD}
A set of optimal points of the block coordinate optimization with respect to ${\bf a}_k$ in (\ref{Prob:Perfectcase_RAO}) is given by
	\begin{align}
		\setcounter{equation}{9} 
	\Big\{ {\bf e}_{m} :     m \in  \mathop{\arg\max}_{n\in\mathcal N} \ Q_{k,n}({\bf a}_{-k}, \epsilon,{\bf p}) \Big\}, \  k\in \mathcal K, \label{PerUpdate_a}
	\end{align}
	and a set of optimal points of the block coordinate optimization with
	respect to $\epsilon$ is given by
	\begin{align}
		\mathop{\arg\max}_{\epsilon\in {{\mathcal B}({\bf a},{\bf p})}\cup\{1\} }
		\bar T ( {\bf A},\epsilon,{\bf p}) .\label{PerUpdate_epsilon}
	\end{align}
	
\end{theorem}
\vspace{-3mm}

\begin{IEEEproof}
First, it is clear that each problem in (\ref{Prob:Perfectcase_RAO}) has the same form as the problem in
\cite[Excersice 4.8]{Boyd2004convex}.
According to the analytical solution of the problem in \cite[Excersice 4.8]{Boyd2004convex}, we can obtain the optimal point of each problem in~(\ref{Prob:Perfectcase_RAO}) as in~(\ref{PerUpdate_a}).
Next, since $\frac{\partial \bar T ({\bf A}, \epsilon,{\bf p})}{\partial \epsilon}$ is a polynomial function of $\epsilon$, we can obtain an optimal point of the problem in (\ref{Prob:Perfectcase_ACB}) by checking all roots of $\frac{\partial \bar T ({\bf A}, \epsilon,{\bf p})}{\partial \epsilon}=0$ and the endpoints of the interval.
Therefore, we can obtain the optimal point of the problem in (\ref{Prob:Perfectcase_ACB}) as in (\ref{PerUpdate_epsilon}).
\end{IEEEproof}

For all $k \in \mathcal K$ and $n \in \mathcal N$, the computational complexity for calculating $Q_{k,n}({\bf a}_{-k}, \epsilon, {\bf p}  ) $ is $\mathcal O ( K2^K)$.
For all $k \in \mathcal K$, the computational complexity for finding the largest one among $ Q_{k,n}({\bf a}_{-k}, \epsilon, {\bf p}  ), n\in \mathcal N$ is $\mathcal O (N)$.
Thus, the overall computational complexity for determining the sets in (\ref{PerUpdate_a}) is $ \mathcal O \left( K \left( N K 2^K + N \right)\right) = \mathcal O \left(  N K^2 2^K\right)$.
In addition,
the roots of equation $ q( {\bf A},z,{\bf p}) =0$ with respect to $z$ can be obtained by solving a
univariate polynomial equation
of degree at most $K-1$,
using mathematical tools, e.g., MATLAB.
The computational complexity for
determining $q({\bf A}, z, {\bf p}  ) $ and $\mathcal B({\bf A},{\bf p})$ are $\mathcal O ( NK2^K)$ and $\mathcal O (K^3)$, respectively.
The computational complexity for computing $\bar T\left( {\bf A}, z,{\bf p}  \right)$, $z \in \mathcal B({\bf A},{\bf p}) \cup \{1\}  $  is $ \mathcal O \left(  NK^2 2^K   \right) $.
The computational complexity for finding the largest ones among $\bar T\left( {\bf A}, z,{\bf p}  \right)$, $z \in \mathcal B({\bf A},{\bf p}) \cup \{1\}  $ is $\mathcal O (K)$.	
The overall computational complexity for determining the set in (\ref{PerUpdate_epsilon}) is $\mathcal O(  NK2^K+K^3 +NK^22^K + K  ) =\mathcal O(  NK^22^K  ) $.
Note that as constants $\sum\nolimits_{{\bf x \in \mathcal X}}
p_{\bf x}
\prod\nolimits_{k \in \mathcal K'}{x}_{k}$, ${\mathcal  K'} \subseteq \mathcal K$ are computed in advance, the corresponding complexities are not considered in the above complexity analysis, and as $\bf A$ is usually sparse during the iterations, the actual computational complexities for obtaining (\ref{PerUpdate_a}) and (\ref{PerUpdate_epsilon}) are much lower.

Based on the proof for \cite[Proposition 2.7.1]{Bertsekas1998NP}, we can prove that Algorithm \ref{Alg_PBCD} returns a stationary point of Problem~\ref{Prob:Perfectcase} in a finite number of iterations.
In practice, we can run Algorithm~\ref{Alg_PBCD} multiple
	times with different feasible initial $\bf A$ to obtain multiple
	stationary points, and choose the stationary point with the largest
	objective value as a suboptimal point of Problem~\ref{Prob:Perfectcase}.

\begin{algorithm}[t]
	\caption{Obtaining A Stationary Point of Problem \ref{Prob:Perfectcase}}
	\label{Alg_PBCD}
		\begin{algorithmic}[1]
		\small{	\STATE \textbf{initialization:} for $k \in \mathcal K$, set ${\bf a}_k := {\bf e}_{n_k}$, where $n_k$ is randomly chosen from $\mathcal N$,
			and set $\epsilon:=1$.
			\STATE \textbf{repeat}
			\STATE  ${\bf A}_{\text{last}} :=\bf A$.
			\STATE \textbf{for $k \in \mathcal K$ do}
			\STATE \ \   \textbf{if} ${\bf a}_k$ does not belong
			             to the set in (\ref{PerUpdate_a})
			\STATE \  \ \ \   ${\bf a}_k $ is randomly chosen
			              from the set in (\ref{PerUpdate_a}).
			\STATE \ \  \textbf{end if}
			\STATE  \textbf{end for}
			\STATE  \textbf{if} $\epsilon$ dose not belong to the set in (\ref{PerUpdate_epsilon})
			\STATE  \ \  $\epsilon$ is randomly chosen
			          from the set in (\ref{PerUpdate_epsilon}).
			\STATE  \textbf{end if}
			\STATE \textbf{until} $  {\bf A}_{\text{last}}=  {\bf A }$.
			}\normalsize
		\end{algorithmic}

\end{algorithm}


\section{Low-complexity Solution }
	\vspace{-2 mm}
From the complexity analysis for obtaining a stationary point of Problem~\ref{Prob:Perfectcase}, we know that
Algorithm~\ref{Alg_PBCD} is computationally expansive when $K$ or $N$ is large.
In this section, we develop another iterative algorithm to obtain a low-complexity solution of Problem~\ref{Prob:Perfectcase}, which is applicable for large $K$ or $N$.
Later, in Section VI, we shall show that such low-complexity algorithm achieves competitive average throughput compared with Algorithm~\ref{Alg_PBCD}, although it has much lower computational complexity than Algorithm~\ref{Alg_PBCD}.

First, we approximate the complicated function $\bar T(\mathbf {A},\epsilon,{\bf p})$, which has $N2^K$ terms, with a simpler function, which has $ 1+ \frac{K(K-1)}{2}$ terms.
Motivated by the approximations of $\bar T(\mathbf {A},\epsilon,{\bf p})$ in \cite{Popovski18SPAWC}, we consider approximate function parameterized by $\lambda \in \mathbb R$ as shown in (\ref{T_LB}) at the top of this page.
Note that $  \sum\nolimits_{{\bf x} \in \mathcal X }p_{\bf x}x_k  $ and $  \sum\nolimits_{{\bf x} \in \mathcal X }p_{\bf x}x_kx_l  $ ($k<l$) represent the probability of device $k$ being active and probability of devices $k$ and $l$ being active, respectively.
Though $  \sum\nolimits_{{\bf x} \in \mathcal X }p_{\bf x}x_k  $ and $  \sum\nolimits_{{\bf x} \in \mathcal X }p_{\bf x}x_kx_l  $ contain $2^K$ terms and are hard to compute, in practice, they can be easily approximated from empirical activity states.
By comparing (\ref{T_LB}) with (\ref{T_Avg.New}), we can see that $ \tilde T_{\lambda}(\mathbf {A},\epsilon,{\bf p})  $ captures the activity probabilities of a single device and every two devices.
For all $( {\bf A }, \epsilon)$ satisfying (\ref{C1}), (\ref{C12}), (\ref{C2}), (\ref{C3}) and all $\bf p$ satisfying (\ref{Cp1}) and (\ref{Cp2}),
we can obtain an upper bound on the approximation error:
\begin{align}
		\setcounter{equation}{12}
	&\!\!\!\!\left| \bar T(\mathbf {A},\epsilon,{\bf p}) - \tilde T_{\lambda}( {\bf A},\epsilon,{\bf p})  \right|
	\leq  \nonumber \\
&\!\!\!\!	2\max(\left| \lambda-1\right|, \left|\lambda\right|  ) \epsilon^2  \!\!
	\sum\limits_{n\in \mathcal {N}} \sum\limits_{k\in \mathcal {K }}\!  a_{k,n} \!\!\!\! \sum\limits_{l \in {\mathcal K}: l >k}  \!\!\!\!   a_{l,n}\!\!
	\sum\limits_{{\bf x} \in \mathcal X }p_{\bf x} x_kx_l  . \label{Errorbound}
\end{align}
The upper bound in (\ref{Errorbound}) is minimized at $\lambda = \frac{1}{2}$, which can be easily shown.
Thus, we approximate $\bar T(\mathbf {A},\epsilon,{\bf p})$ with $\tilde T_{\frac{1}{2}}\left(\mathbf {A},\epsilon,{\bf p}\right)$, and consider the following approximate problem of Problem~\ref{Prob:Perfectcase}.

\begin{Prob}
 	[Approximate Average Throughput Maximization]
 \vspace{-3mm}
	\label{Prob:Perfectcase_LB}
	{\begin{align}
		&{\max_{\bf{A},\epsilon}} \quad  \tilde T_{\frac{1}{2}}\left(\mathbf {A},\epsilon,{\bf p}\right) \nonumber\\
		& \ \mathrm{s.t.} \quad \text{(\ref{C1})},\  \text{(\ref{C12})},\ \text{(\ref{C2})},  \ \text{(\ref{C3})} .\nonumber
	\end{align}}
\end{Prob}

Analogously, using the BCD method, we propose a computationally efficient iterative algorithm to obtain a stationary point of Problem \ref{Prob:Perfectcase_LB}, which has more performance guarantee than the heuristic method in \cite{Popovski18SPAWC}.
Specifically, variables $\left({\bf A}, \epsilon\right)$ are divided into $K+1$ blocks, i.e., $ {\bf a}_{k}, \ k \in \mathcal K $ and $ \epsilon $.
For ease of illustration, in the following, we also write $\tilde T_{\frac{1}{2}}(\mathbf {A},\epsilon,\bf p)$ as $\tilde T_{\frac{1}{2}}( {\bf a}_{k},{\bf a}_{-k} ,\epsilon,\bf p)$.
Given ${\bf a}_{-k}$ and $\epsilon$ obtained in the previous step, the block coordinate optimization with respect to ${\bf a}_{k}$ is
given by
\begin{align}
	&\max_{{\bf a}_k} \quad
	\tilde T_{\frac{1}{2}}\left( {\bf a}_k ,{\bf a}_{-k} ,   \epsilon,{\bf p}\right)
	,\quad k \in \mathcal K\label{Prob:Perfectcase_RAOLB}\\
	&\ \mathrm{s.t.}   \quad   \text{(\ref{C2})}, \ \text{(\ref{C3})}.  \nonumber
\end{align}
Given $\bf A$ obtained in the previous step, the block coordinate optimization with respect to $\epsilon$ is given by
\begin{align}
	&\max_{\epsilon} \quad
	\tilde T_{\frac{1}{2}} \left({\bf A},\epsilon ,{\bf p}  \right)
	\label{Prob:Perfectcase_ACBLB}  \\
	&  \ \mathrm{s.t.}  \quad \      \text{(\ref{C1})}, \ \text{(\ref{C12})}. \nonumber
\end{align}
Each problem in (\ref{Prob:Perfectcase_RAOLB}) is an LP with $N$ variables and $N+1$ constraints, and the problem in (\ref{Prob:Perfectcase_ACBLB}) is a quadratic program (QP) with a single variable and two constraints.
It is clear that the convex problems in (\ref{Prob:Perfectcase_RAOLB}) and (\ref{Prob:Perfectcase_ACBLB}) are much simpler than those in
(\ref{Prob:Perfectcase_RAO}) and (\ref{Prob:Perfectcase_ACB}), respectively.

Based on structural properties of the block coordinate optimization problems in (\ref{Prob:Perfectcase_RAOLB}) and (\ref{Prob:Perfectcase_ACBLB}), we can obtain their optimal points.

\begin{theorem}[Optimal Points of Problems in (\ref{Prob:Perfectcase_RAOLB}) and (\ref{Prob:Perfectcase_ACBLB})]\label{Thm_OptimmalSolutionBCDLB}
A set of optimal points of the block coordinate optimization
	with respect to ${\bf a}_k$ in (\ref{Prob:Perfectcase_RAOLB}) is given by
	\begin{align}
		\Big\{ {\bf e}_{m} :     m \in  \mathop{\arg\min}_{n\in\mathcal N}
		\sum\limits_{l \in {\mathcal K}: l \neq k}   a_{l,n}\sum\limits_{{\bf x} \in \mathcal X }p_{\bf x} x_k x_l \Big\}
		,\ k\in\mathcal K,  \label{PerLBUpdate_a}
	\end{align}
	and the optimal point of the block coordinate optimization with
	respect to $\epsilon$ is given by
	\begin{align}
		\min\left(1,  \frac{\sum\limits_{k\in \mathcal {K}}
			\sum\limits_{{\bf x} \in \mathcal X }p_{\bf x}x_k}
		{
			2\sum\limits_{n\in \mathcal {N}} \sum\limits_{k\in \mathcal {K}}a_{k,n}\sum\limits_{l \in {\mathcal K}:l> k} a_{l,n} \sum\limits_{{\bf x} \in \mathcal X }p_{\bf x} x_kx_l   }\right ).\label{PerLBUpdate_epsilon}
	\end{align}

\end{theorem}

\begin{IEEEproof}
	Theorem \ref{Thm_OptimmalSolutionBCDLB} can be proved in a similar way to Theorem \ref{Thm_OptimmalSolutionBCD}. We omit the details due to
	page limitation.
\end{IEEEproof}

For all $k \in \mathcal K$ and $n \in \mathcal N$, the computational complexity for
calculating $\sum\nolimits_{l \in {\mathcal K}: l \neq k}   a_{l,n}\sum\nolimits_{{\bf x} \in \mathcal X }p_{\bf x} x_k x_l$ is $\mathcal O ( K)$.
For all $n \in \mathcal N$, the computational complexity of finding the largest one among $ \sum\nolimits_{l \in {\mathcal K}: l \neq k}   a_{l,n}\sum\nolimits_{{\bf x} \in \mathcal X }p_{\bf x} x_k x_l$, $n\in \mathcal N$ is $\mathcal O (N)$.
Thus, the overall computational complexity for determining the sets in (\ref{PerLBUpdate_a}) is $\mathcal O (  K(K+N)) =  \mathcal O (  NK+K^2) $.
The computational complexity for obtaining the closed-form optimal point in~(\ref{PerLBUpdate_epsilon}) is $\mathcal O(NK^2)$.
Note that constants $\sum\nolimits_{k\in \mathcal {K}}
\sum\nolimits_{{\bf x} \in \mathcal X }p_{\bf x}x_k$ and $\sum\nolimits_{{\bf x} \in \mathcal X }p_{\bf x}x_lx_k $, $k,l \in \mathcal K,k<l $ are computed in advance,
and hence the corresponding complexities are not considered in the above complexity analysis.
It is obvious that the computational complexities for obtaining the optimal points given by Theorem~\ref{Thm_OptimmalSolutionBCDLB} are much lower than those for obtaining the optimal points given by Theorem~\ref{Thm_OptimmalSolutionBCD}.
Furthermore, it is worth noting that the optimal points
given by Theorem~\ref{Thm_OptimmalSolutionBCDLB} do not rely on the activity correlation of more than two devices.

Based on the proof for \cite[Proposition 2.7.1]{Bertsekas1998NP}, we can prove that Algorithm \ref{Alg_PLBBCD} returns a stationary point of Problem~\ref{Prob:Perfectcase_LB} in a finite number of iterations.
Similarly, we can run Algorithm~\ref{Alg_PLBBCD} multiple
	times with different feasible initial $\bf A$ to obtain multiple
	stationary points of Problem~\ref{Prob:Perfectcase_LB}, and choose the stationary point with the largest average throughput as a suboptimal point of Problem~\ref{Prob:Perfectcase}.

\begin{algorithm}[t]
	\caption{Obtaining A Stationary Point of Problem \ref{Prob:Perfectcase_LB}}
	\label{Alg_PLBBCD}
		\begin{algorithmic}[1]
			\small{\STATE \textbf{initialization:}
			for $k \in \mathcal K$, set ${\bf a}_k := {\bf e}_{n_k}$, where $n_k$ is randomly chosen from $\mathcal N$,
			and set $\epsilon:=1$.
			\STATE \textbf{repeat}
			\STATE ${\bf A}_{\text{last}}:=\bf A$.
			\STATE \textbf{for $k \in \mathcal K$ do}
			\STATE \ \  \textbf{if} ${\bf a}_k$ does not belong
			           to the set in  (\ref{PerLBUpdate_a})
			\STATE \  \ \ \  ${\bf a}_k $ is randomly chosen
			          from the set in  (\ref{PerLBUpdate_a}).
			\STATE \ \  \textbf{end if}
			\STATE  \textbf{end for}
			\STATE  \textbf{if} $\epsilon$ dose not belong to the set in (\ref{PerLBUpdate_epsilon})
			\STATE  \ \  $\epsilon$ is randomly chosen
			         from the set in  (\ref{PerLBUpdate_epsilon}).
			\STATE  \textbf{end if}
			\STATE  \textbf{until} $  {\bf A}_{\text{last}}=  {\bf A }$.
			}\normalsize
		\end{algorithmic}
\end{algorithm}

	\vspace{-3mm}

\section{Numerical Results}
	\vspace{-1mm}
In this section, we evaluate the performance of the proposed
solutions via numerical results.
We consider three baseline schemes, namely BL-MMPC, BL-MSPC and BL-LTE.
In BL-MMPC and BL-MSPC, $ {\bf a}_k$, $k\in \mathcal K$ are obtained by the MMPC and MSPC allocation algorithms in \cite{Popovski18SPAWC}, respectively, and $\epsilon =1$.
In BL-LTE, we set $a_{k,n} = \frac{1}{N}, \ k\in\mathcal K, n\in\mathcal N$ according to the standard random access procedure of LTE networks~\cite{TR_3gppevolved}, and set $\epsilon= \min \left(1,\frac{N}{\bar K}\right)$ according to the optimal access control~\cite{Wong16TVT}, where $\bar K$ denotes the average number of active devices. 
Note that  BL-MMPC and BL-MSPC make use of the correlation of the activities of every two devices; BL-LTE does not utilize any information on correlation of device activities.
In the simulation, we adopt the group device activity model.
Specifically, $K$ devices are divided into $G$ groups each of size $ \frac{K}{G}$ (assuming $K$ is divisible by $G$),
the activity states of devices in different groups are independent,
and the activity states of devices in a group are the same.
The probability that a group is active, i.e., all devices in this group are active, is
$p_a$.
Then, $\bf x$ and $p_{\bf x}, {\bf x}\in \mathcal X$ can be easily
determined.

\begin{figure}[t]
	\begin{center}
		\subfigure[\footnotesize{Average throughput versus $K$ at $N=15$. }]
		{\resizebox{5.2 cm}{!}{\includegraphics{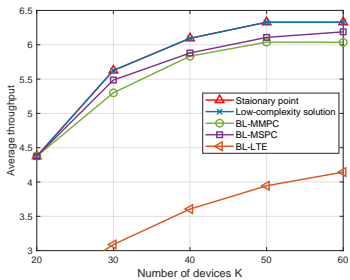}}}   \\
		\vspace{-3.5mm}
		\subfigure[\footnotesize{Average throughput versus $N$ at $K=60$.}]
		{\resizebox{5.2 cm}{!}{\includegraphics{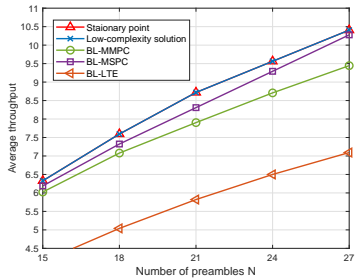}}}
	\end{center}
	\vspace{-6mm}
	\caption{\small{Average throughput comparision between the proposed solutions and baseline schemes at $\frac{K}{G} =10$, $\varepsilon = 0.3$ and $p_a=0.25$.
	}}
	\vspace{-6mm}
	\label{Figure_Throughput_small_K}
	
\end{figure}

First, we compare the average throughput of the proposed solutions and three baseline schemes, at small numbers of devices and preambles.
Fig.~\ref{Figure_Throughput_small_K} illustrates the average throughput versus
the number of devices $K$ and the number of preambles $N$.
From Fig.~\ref{Figure_Throughput_small_K}, we make the following observations.
The proposed stationary point significantly outperforms BL-MMPC and BL-MSPC, as the stationary point relies on $p_{\bf x} $, ${\bf x} \in \mathcal X$, which capture the correlation of the activities of all devices;
BL-MMPC and BL-MSPC outperform BL-LTE, as BL-MMPC and BL-MSPC both make use of the correlation of the activities of every two devices.
The proposed low-complexity solution outperforms BL-MMPC and BL-MSPC, as it relies on a more accurate approximation of the average throughput and is obtained by a more effective algorithm;
it is worth noting that the gap between the average throughput of the stationary point and the low-complexity solution is small,
which shows that exploiting the correlation of the activities of every two devices in a rigorous way already achieves a significant gain.
Furthermore, from Fig.~\ref{Figure_Throughput_small_K}~(a), we can see that the average throughput of each scheme increases with $K$, due to the increase of traffic load.
From Fig.~\ref{Figure_Throughput_small_K}~(b), we can see that the average throughput of each scheme increases with $N$, due to the increase of communications resource.

\begin{figure}[t]
	\begin{center}
		\subfigure[\footnotesize{Average throughput versus $K$ at $N =50$.}]
		{\resizebox{5.2  cm}{!}{\includegraphics{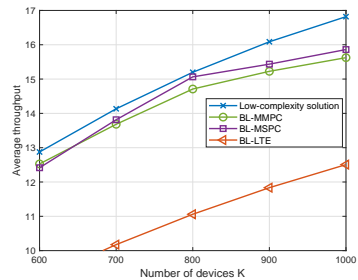}}}
			\\	\vspace{-3.5mm}
		\subfigure[\footnotesize{Average throughput versus $N$ at $K = 1000$. }]
		{\resizebox{5.2 cm}{!}{\includegraphics{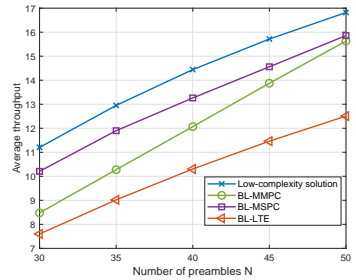}}}
	\end{center}
	\vspace{-5mm}
	\caption{\small{Average throughput comparision between the proposed solutions and baseline schemes at $\frac{K}{G}  = 20$, $\varepsilon = 0.3$ and $p_a = 0.03$.
	}}
	\vspace{-6mm}
	\label{Figure:TotalUtility}
\end{figure}

Next, we compare the average throughputs of the proposed solutions and three baseline schemes, at large numbers of devices and preambles.
Fig.~\ref{Figure:TotalUtility} illustrates the average throughput versus
the number of devices $K$ and the number of preambles $N$.
From Fig.~\ref{Figure:TotalUtility}, we also observe that the low-complexity solution  significantly outperforms BL-MMPC, BL-MSPC and BL-LTE;
the results at large $K$ and $N$ shown in Fig.~\ref{Figure:TotalUtility} are similar to those at small $K$ and $N$ shown in Fig.~\ref{Figure_Throughput_small_K}.

\vspace{-3 mm}
\section{Conclusion}
\vspace{-1mm}
In this paper, we investigated the joint optimization of preamble selection and access barring for correlated device activities which exist in most IoT applications and is important for 6G.
We optimized the preamble selection
distribution and the access barring factor to maximize the average throughput of the devices.
We characterized an optimality property and
obtained a stationary point and a low-complexity
solution.
We numerically showed that the two proposed
solutions achieve significant gains over existing schemes and have a small gap in average throughput.
The numerical results demonstrate the significance of exploiting the correlation of activities of every two devices
in a rigorous way.
Both the theoretical and numerical results offer important design insights for MTC.

\vspace{-1mm}

\maketitle




\IEEEpeerreviewmaketitle

\ifCLASSOPTIONcaptionsoff
  \newpage
\fi





\end{document}